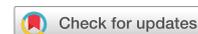

ARTICLE   OPEN

# Computing grain boundary diagrams of thermodynamic and mechanical properties

Chongze Hu [1,2], Yanwen Li[3], Zhiyang Yu[3,4] and Jian Luo [1,2 ✉]

Computing the grain boundary (GB) counterparts to bulk phase diagrams represents an emerging research direction. Using a classical embrittlement model system Ga-doped Al alloy, this study demonstrates the feasibility of computing temperature- and composition-dependent GB diagrams to represent not only equilibrium thermodynamic and structural characters, but also mechanical properties. Specifically, hybrid Monte Carlo and molecular dynamics (MC/MD) simulations are used to obtain the equilibrium GB structure as a function of temperature and composition. Simulated GB structures are validated by aberration-corrected scanning transmission electron microscopy. Subsequently, MD tensile tests are performed on the simulated equilibrium GB structures. GB diagrams are computed for not only GB adsorption and structural disorder, but also interfacial structural and chemical widths, MD ultimate tensile strength, and MD tensile toughness. This study suggests a research direction to investigate GB composition–structure–property relationships via computing GB diagrams of thermodynamic, structural, and mechanical (or potentially other) properties.

npj Computational Materials (2021)7:159 ; https://doi.org/10.1038/s41524-021-00625-2

## INTRODUCTION

In polycrystalline materials, grain boundaries (GBs) are the ubiquitous crystal imperfection that can often control materials fabrication processing and performance[1–6]. It has been long recognized that GBs can be treated as two-dimensional (2D) interfacial phases[7], which is more recently termed as "complexions" to differentiate them from thin GB precipitation layers of bulk (3D) phases[1,2,8–10]. Recently, it was proposed to develop the GB counterparts to bulk phase diagrams as a generally useful materials science tool[11]. Several types of GB diagrams have been computed via thermodynamic models[11–19], density functional theory calculations[20], atomistic simulations[21–23], and machine learning[21]. To date, GB diagrams have only been developed to represent a limited number of thermodynamic and structural properties, mostly notably adsorption (i.e., the GB excess of solutes)[18,21–24] and interfacial structural disorder[11,14–17,21–24]. This work strives for further constructing GB diagrams to represent not only other useful equilibrium structural characters (e.g., GB structural and chemical widths) but also computed mechanical properties.

Solute or impurity segregation (a.k.a. adsorption in interfacial thermodynamics) at GBs can substantially alter their structural characters to influence various mechanical and other functional properties[1,3–5,25–31]. Specifically, GB embrittlement (GBE) induced by segregation is one of the most classical phenomena in physical metallurgy[4,26,28–34]. While the Rice–Wang model offers a general thermodynamic framework[31], the atomistic mechanisms underpinning GBE are not fully understood, defying scrutiny over a century. Liquid metal embrittlement (LME), where normally ductile metals can fail catastrophically in contact with certain liquid metals, represents a particular mystery in materials science[35–37]. While there are other contributing factors, LME often occurs along with severe GBE. Here, Ni–Bi[28] and Al–Ga[32–34,38–43] represent two classical systems that exhibit both severe GBE and LME.

This study aims at using a classical GBE and LME system, Al–Ga (Ga-doped Al), as a model system to establish an exemplar to investigate GB composition–structure–property relationships via computing GB diagrams of thermodynamic, structural, and mechanical properties.

Moreover, the majority of prior atomistic studies focused on symmetric tilt or twist GBs that are relatively easy to image and model. The atomic-level embrittlement mechanisms of general GBs (a.k.a. asymmetric GBs that are often of mixed tilt and twist characters) are poorly understood. However, the asymmetric and general GBs are ubiquitous and can often be the weaker link mechanically and chemically in polycrystalline materials that limit the performance. Recent studies of two classical GBE systems, Bi vs. S doped Ni, revealed different characteristic interfacial structures that underpin the embrittlement of general GBs. In Ni–Bi, ordered bilayer adsorption of Bi caused LME[28], where each of the adsorbed layers can undergo reconstruction to form highly ordered interfacial superstructures[28]. In Ni–S, more disordered bilayer-like and amorphous-like segregation structures with bipolar characters have been discovered[26]. However, the lack of fast and reliable interatomic potentials in these systems prevents us from computing realistic GB diagrams via large-scale atomistic simulations. Al–Ga represents another classical GBE and LME system that exhibits different thermodynamic characters (e.g., Ga has substantial solid solubility in Al vs. limited solid solubilities of Bi or S in Ni) and interfacial structures[32–34,38–41,43]. A fast and reliable embedded-atom model (EAM) potential has been developed and tested for Al–Ga[34,40,44], which enables us to use Al–Ga as our model system to compute GB diagrams to establish an exemplar. We should note that the bulk liquid Ga can also wet and penetrate to Al GBs to contribute to LME[38,41,45]. The current study, however, focuses on the (undersaturated) single bulk solid solution phase region where severe GBE occurs (before the appearance of the bulk Ga-based liquid metal).

[1]Department of Nanoengineering, University of California San Diego, La Jolla, CA 92093, USA. [2]Program of Materials Science and Engineering, University of California San Diego, La Jolla, CA 92093, USA. [3]State Key Laboratory of Photocatalysis on Energy and Environment, College of Chemistry, Fuzhou University, 350002 Fuzhou, Fujian, P.R. China. [4]Department of Physics, Southern University of Science and Technology, Shenzhen, Guangdong 518055, P. R. China. ✉email: jluo@alum.mit.edu

Using a general GB in Ga-doped Al as the model system, we simulated equilibrium GB structure as a function of temperature and bulk Ga fraction. Subsequently, we conducted molecular dynamics (MD) tensile tests to study the mechanical response of the equilibrium GB structure with external loading. Consequently, we construct GB diagrams for not only Ga adsorption and interfacial disorder (that have been computed only for a limited number of three other systems[21–23]) but also interfacial structural and chemical widths and computed mechanical properties (for the first time). This study expands our capability to compute GB diagrams via atomistic simulations.

## RESULTS
### Validation of trilayer-like segregation in a symmetric GB
We adopted isobaric semi-grand canonical (constant $N(\Delta\mu)PT$) ensemble hybrid Monte Carlo and molecular dynamics (MC/MD) simulations to obtain equilibrium GB structures in Ga-doped Al using an EAM potential originally developed by the Srolovitz group[44]. The EAM potential can reproduce Al–Ga binary phase diagram in agreement with experiments (as shown in Supplementary Fig. 1). See the "Methods" section for the detailed computation procedure.

To validate our simulation methods and the EAM potential, we first model the trilayer-like segregation structure at a symmetric tilt Σ11 GB reported by Sigle et al. (Fig. 1a)[33]. This largely ordered trilayer-like structure also serves as a relatively simple starting example to investigate Ga segregation at Al GBs. Here, we constructed this Σ11 (113)//(113) GB structure in a unit cell with 18480 atoms to perform hybrid MC/MD simulations. Subsequently, we simulated the scanning transmission electron microscopy (STEM) image based on the simulated GB structure (Fig. 1c). The simulated STEM image clearly shows a trilayer-like Ga segregation structure (Fig. 1b), in good agreement with the prior experimental high-resolution TEM (HRTEM) image (Fig. 1a)[33]. Furthermore, the 2D averaged Ga atomic fraction profile ($X_{Ga}$) across the GB (Fig. 1f) shows a triple peaks pattern at the GB, consistent with the experimental observation.

In addition, we calculated the structural disorder parameter ($\eta_{Dis}$) from the simulated GB structure for each atom following a procedure proposed by Chua et al. (where $\eta_{Dis} = 0$ for a perfect crystal and $\eta_{Dis} = 1$ for a liquid)[46]. A corresponding color map of $\eta_{Dis}$ is shown in Fig. 1d and the 2D averaged $\eta_{Dis}$ profile is shown in Fig. 1g. The GB core has a maximal $\eta_{Dis}$ value of ~0.5, which suggests the structure of this symmetric tilt GB to be relatively ordered (in comparison with the asymmetric and general GB shown in Fig. 2 and discussed in the next section). Moreover, we calculated centrosymmetry parameters (CSP) to measure the local lattice disorder using the OVITO code (where CSP = 0 represents the ideal crystal and positive CSP values indicate defects; see the "Methods" section)[47,48]. The CSP map (Fig. 1e) and the 2D averaged CSP profile (Fig. 1h) again confirmed the relatively ordered segregation structure at this symmetric tilt GB with a small maximal CSP value of ~7 (vs. ~18 for the asymmetric and general GB shown in Fig. 2). In summary, our hybrid MC/MD simulations can reproduce the trilayer-like Ga segregation structure in the Al Σ11 symmetric tilt GB that was previously observed in a bicrystal experiment[33]. We have further revealed structural details of this (relatively simple and ordered) symmetric tilt GB.

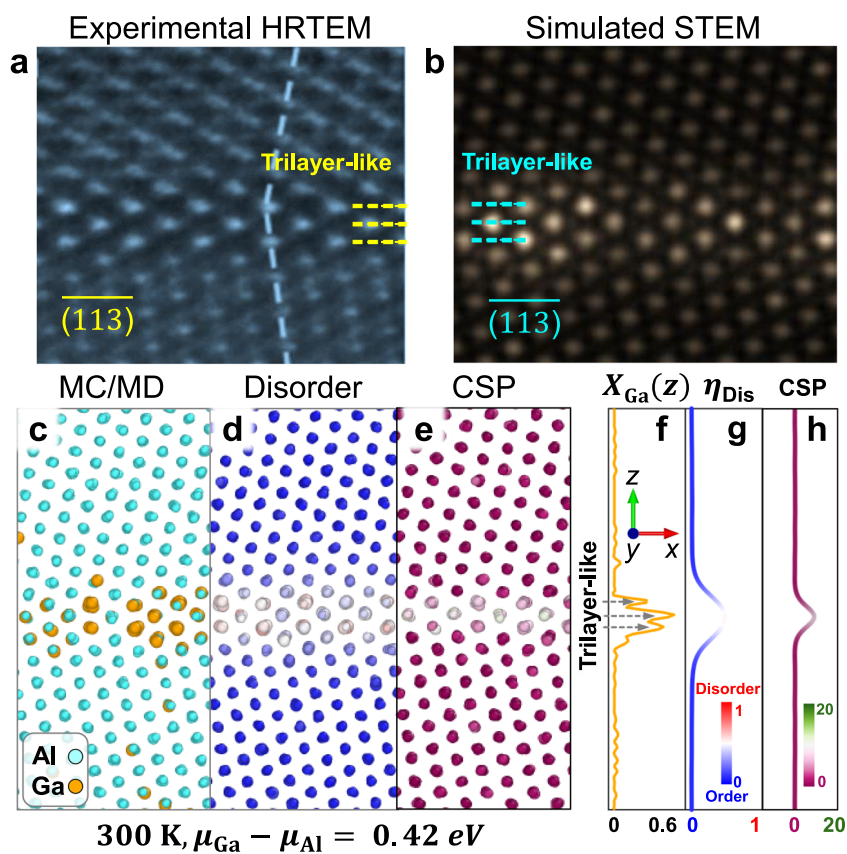

**Fig. 1 Validation of the computational approach with ordered segregation at a symmetric tilt GB. a** HRTEM of a Ga-doped Al symmetric tilt Σ11 (113)//(113) GB, reprinted from ref. [33]. with permission. **b** Simulated STEM images of the same symmetric tilt GB based on the equilibrium atomistic structure obtained from hybrid MC/MD simulations at 300 K and $\mu_{Ga} - \mu_{Al} (\equiv -\Delta\mu) = 0.42$ eV. **c** The hybrid MC/MD simulated GB structure and the corresponding color maps based on **d** computed disorder parameter ($\eta_{Dis}$) and **e** centrosymmetric parameter (CSP) on each atom. The 2D averaged **f** Ga atomic fraction ($X_{Ga}(z)$; $X_{Ga}(\pm\infty) = X$), **g** disorder $\eta_{Dis}(z)$, and **h** CSP profiles across the GB as functions of the spatial variable $z$ perpendicular to the GB.

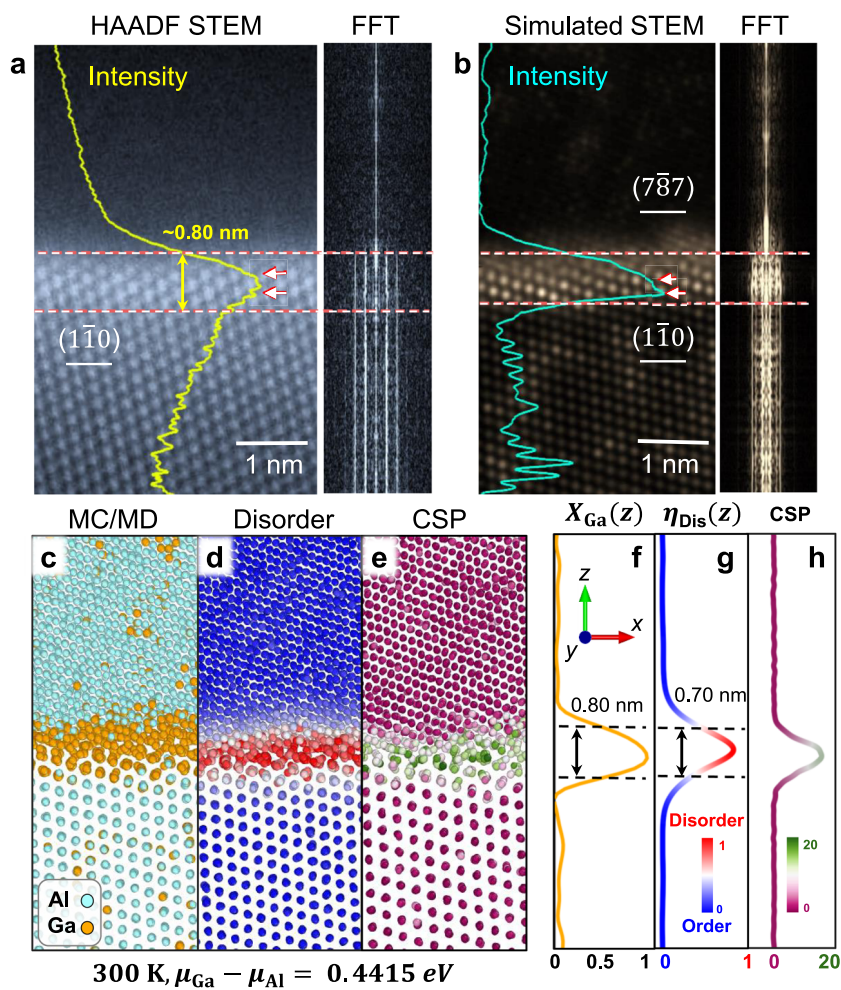

**Fig. 2 The Ga segregation structure in an asymmetric general Al GB. a** HAADF STEM of a general GB randomly selected from a Ga-doped Al polycrystalline specimen, where the orientation of the low-index grain surface is $(1\bar{1}0)$. **b** Simulated STEM images of an asymmetric Σ81 $(1\bar{1}0)//(7\bar{8}7)$ GB based on the equilibrium atomistic structure obtained from hybrid MC/MD simulations at 300 K and $\mu_{Ga} - \mu_{Al} (\equiv -\Delta\mu) = 0.4415$ eV. **c** The hybrid MC/MD simulated GB structure and the corresponding color maps based on the **d** computed disorder parameter ($\eta_{Dis}$) and **e** centrosymmetric parameter on each atom. The 2D averaged **f** Ga atomic fraction $X_{Ga}(z)$, **g** disorder $\eta_{Dis}(z)$, and **h** CSP profiles across the GB as functions of the spatial variable $z$ perpendicular to the GB. In STEM images, only the order of one grain in the low zone axis is visible, while the other (off-aligned) grain should also impose partial order to the segregation zone with a more disordered GB core in between. This "graded order/disorder/order" segregation structure at 300 K is schematically illustrated in Supplementary Fig. 2e.

### Segregation in an asymmetric general GB

Next, we investigated more complex and disordered segregation structures at asymmetric and general GBs. The detailed procedure for our polycrystal experiment was described in the "Methods" section. Figure 2a shows an aberration-corrected (AC) STEM high-angle annular dark-field (HAADF) image of a representative asymmetric and general GB randomly selected from a Ga-doped Al polycrystalline specimen, where the orientation of the low-index grain surface is $(1\bar{1}0)$. The high brightness at the GB region suggests strong Ga segregation (Fig. 2a), which is evident quantitatively in the intensity profile (the yellow line in Fig. 2a). Moreover, the intensity profile shows partial order in the GB region (indicated by arrows in Fig. 2a). The partially ordered Ga segregation structure can be further analyzed and demonstrated by performing line-by-line fast Fourier transformation (FFT) analysis of the STEM image, which is shown in the right panel of Fig. 2a.

To mimic this experimentally observed general GB shown in Fig. 2a, we selected an asymmetric Σ81 $(1\bar{1}0)//(7\bar{8}7)$ GB of mixed tilt and twist characters for simulations. Here, one grain has the $(1\bar{1}0)$ terminal surface matching the general GB randomly selected in a polycrystal shown in Fig. 2a. The orientation of the other grain was selected to maintain a required periodic boundary condition while having a large Σ value of 81 to best represent a general GB. Subsequently, we performed a hybrid MC/MD simulation at 300 K and $\Delta\mu = -0.4415$ eV (selected so that the simulated interfacial width matched the experimental observation, which corresponds to $X = \sim 0.28 X_{Max}$). The hybrid MC/MD simulated GB structure was shown in Fig. 2c, which clearly shows strong Ga segregation at the GB region. Based on this simulated GB structure, we further performed STEM image simulation (Fig. 2b). The simulated STEM image agrees well with the experimental STEM image (Fig. 2b vs. 2a). The intensity peak and FFT analysis again confirm the partially ordered Ga segregation structure, which becomes more disordered at the GB core (the most middle region between the two abutting grains). To better quantify the structural disorder, we calculated the $\eta_{Dis}$ parameter for each atom from the simulated GB structure. The color map of the $\eta_{Dis}$ parameter (Fig. 2d) shows that the GB core is highly disordered with a $\eta_{Dis}$ value of $\sim$1 (liquid-like), but Ga atoms close to bulk phases are more ordered with smaller $\eta_{Dis}$. The 2D averaged structural disorder profile $\eta_{Dis}(z)$ is shown in Fig. 2g. In addition, the GB core region has a large maximal CSP value of ~18 (Fig. 2h), which is ~2.5× higher than that in the ordered (trilayer-like) symmetric tilt Σ11 GB (Fig. 1h). The combination of the $\eta_{Dis}$ and CSP profiles, along with the experimental and simulated STEM images, clearly confirms the (more) disordered Ga segregation structure at the asymmetric Σ81 GB.

By calculating the full widths at half maximum (FWHMs) of the Ga compositional profile $X_{Ga}(z)$ and structural disorder profile $\eta_{Dis}(z)$, we found that the GB (disordered) structural width ($\delta_{GB}^{Strcuratal}$) calculated from the $\eta_{Dis}(z)$ profile is ~0.70 nm, which is smaller than the GB chemical width ($\delta_{GB}^{Chemical}$) of ~0.80 nm calculated from the $X_{Ga}(z)$ profile. Note that we used the same coarse-graining scheme (a step width of ~0.12 nm and sampling number of 150) to calculate $X_{Ga}(z)$ and $\eta_{Dis}(Z)$ profiles so that the FWHMs ($\delta_{GB}^{Chemical}$ and $\delta_{GB}^{Strcuratal}$) can be compared. This suggests that the Ga atoms near bulk regions are more ordered than those in the disordered GB core region at this low temperature (e.g., 300 K).

In summary, our hybrid MC/MD simulations can reproduce not only the ordered trilayer-like segregation structure in the (special) symmetric tilt GB but also a more disordered segregation structure in more general and asymmetric GB observed in experiments. Additional examples of asymmetric GBs in Al–Ga were reported in a most recent publication[43], which suggested similar (disorder vs. order gradients) characters of the Ga segregation structures among different general GBs. In addition, the MC/MD-simulated Al–Ga binary phase diagram agrees well with the experimental phase diagram (Supplementary Fig. 1). Therefore, our hybrid MC/MD simulations are validated. Since our primary interest is on the general GBs that are the weak link for mechanical and chemical degradation, we will use the asymmetric Σ81 GB to represent general GBs to compute GB diagrams of both thermodynamic and mechanical properties in the subsequent in-depth investigation.

### GB diagrams of thermodynamic and structural properties

To compute GB diagrams, we performed a series of isobaric semi-grand canonical ensemble hybrid MC/MD simulations to compute the equilibrium GB structure as a function of temperature ($T$) and chemical potential difference ($\Delta\mu$). See the "Methods" section for details of the simulations. Via hybrid MC/MD simulations of the bulk phases (grains), we also established a relation between $\Delta\mu$ and the bulk Ga fraction ($X \equiv X_{Ga}$) at each temperature. Subsequently, we can compute various GB properties (based on the simulated equilibrium GB structures) and plot the properties as functions of normalized temperature ($T/T_M$, where $T_M$ is the melting temperature of Al) and normalized bulk Ga fraction ($X/X_{Max}$, where $X_{Max}$ is the maximal solid solubility of Ga in Al) to construct GB diagrams.

We first computed two basic thermodynamic properties, GB adsorption (i.e., the GB excess of Ga, $\Gamma_{Ga}$) and GB excess of structural disorder ($\Gamma_{Disorder}$), as functions of $T/T_M$ and $X/X_{Max}$, via the following procedures. The GB excess of Ga can be determined by

$$\Gamma_{Ga} = \int_{-\infty}^{+\infty} [X_{Ga}(z) - X_{Ga}(\pm\infty)] dz,$$

where $X_{Ga}(z)$ is the 2D averaged Ga fraction profile (see, e.g., Fig. 2f) and $X_{Ga}(\pm\infty) \equiv X$ is the bulk Ga fraction. The GB excess of the structural disorder can be computed via

$$\Gamma_{Disorder} = \int_{-\infty}^{+\infty} [\eta_{Dis}(z) - \eta_{Dis}(\pm\infty)] dz,$$

where $\eta_{Dis}(z)$ is the 2D averaged disorder parameter profile (see, e.g., Fig. 2g) and $\eta_{Dis}(\pm\infty) \approx 0$ inside a crystalline grain ($\eta_{Dis} = 0$ for a perfect crystal at 0 K). The computed GB adsorption and structural disorder diagrams are plotted in Fig. 3a and b, respectively.

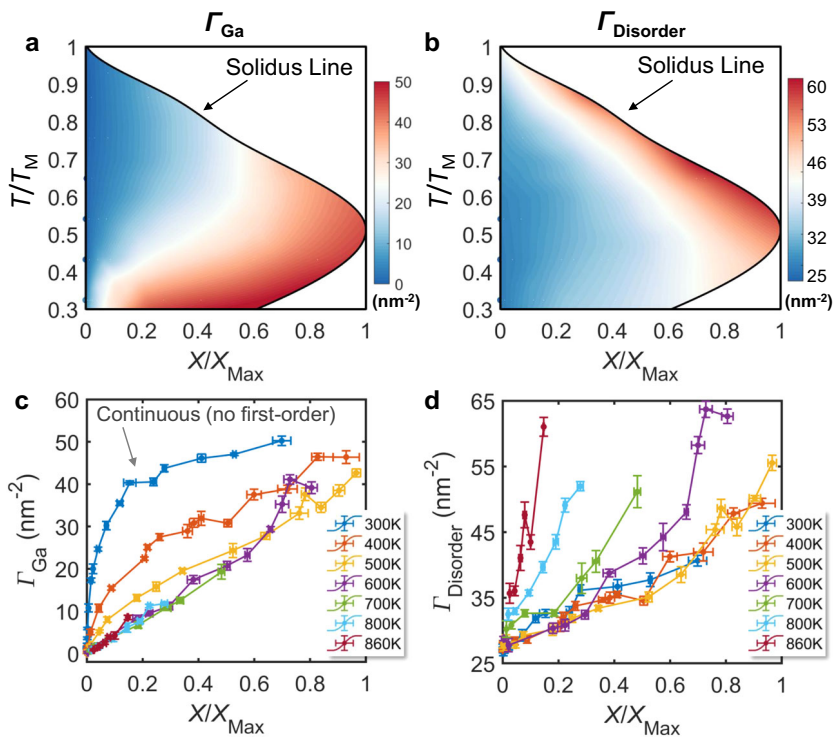

**Fig. 3 Computed GB diagrams of adsorption and structural disorder.** The hybrid MC/MD simulated diagrams of **a** adsorption or GB excess of Ga ($\Gamma_{Ga}$) and **b** GB excess of structural disorder ($\Gamma_{Disorder}$) of the Ga-doped Al asymmetric Σ81 GB. Computed (**c**) $\Gamma_{Ga}$ vs. normalized bulk composition ($X/X_{Max}$) and **d** $\Gamma_{Disorder}$ vs. $X/X_{Max}$ curves at various temperatures. The error bars for $\Gamma_{Ga}$, $\Gamma_{Disorder}$, and $X/X_{Max}$ are the standard derivations of these parameters that are calculated from five different structures during the final stage of hybrid MC/MD simulations at each temperature ($T$) and chemical potential difference ($\Delta\mu$).

Figure 3a shows that the GB adsorption is high at low temperatures, which increases with increasing bulk Ga fraction. The corresponding $\Gamma_{Ga}$ vs. $X/X_{Max}$ curves at different temperatures plotted in Fig. 3c show continuous changes without a first-order adsorption transition, which can be ascribed to an attractive Al–Ga pair-interaction illustrated by a phenomenological model (that will be discussed subsequently). The computed GB structural disorder diagram (Fig. 3b), as well as the corresponding $\Gamma_{Disorder}$ vs. $X/X_{Max}$ curves (Fig. 3d), illustrates that the GB excess structural disorder $\Gamma_{Disorder}$ gradually increases with increasing temperature and bulk Ga fraction. Figure 3b also suggests that Ga addition can induce more structural disorder near the upper segment of the bulk solidus line (for $T > \sim 0.4T_m$).

Such GB adsorption and structural disorder diagrams have been constructed (only) for three other systems, Mo–Ni[22], Si–Au[23], and Cu–Ag[21], in prior studies (and only for the last case of Cu–Ag for asymmetric and general GBs in a most recent study[21]). Here, Al–Ga represents the first GBE/LME system, for which GB adsorption and structural disorder diagrams have been constructed for a general GB via atomistic simulations. In comparison with the symmetric tilt Σ5 GB in Mo–Ni reported in a prior study[22], there is significantly less coupling between the GB adsorption and structural disorder in this asymmetric Σ81 mixed GB in Al–Ga (see Fig. 3a vs. 3b). To establish and understand the GB composition–structure–property relationship, here we have taken a step forward to further compute GB diagrams to represent other GB equilibrium structural characters, as well as simulated mechanical properties in the next section, for the first time to our knowledge.

Specifically, the effective GB chemical and structural widths ($\delta_{GB}^{Chemical}$ and $\delta_{GB}^{Structural}$) were calculated based on FWHMs of the 2D averaged composition profile $X_{Ga}(z)$ and structural disorder profiles $\eta_{Dis}(z)$ (see, e.g., Fig. 2f and h). The computed GB diagrams of $\delta_{GB}^{Chemical}$ and $\delta_{GB}^{Structural}$ were plotted in Fig. 4a and b, respectively. Here, we used the same coarse-graining scheme to calculate $X_{Ga}(z)$ and $\eta_{Dis}(z)$ profiles so that the computed FWHMs ($\delta_{GB}^{Chemical}$ and $\delta_{GB}^{Strcuratal}$) are more (albeit still not rigorously) comparable.

On one hand, the computed GB chemical width ($\delta_{GB}^{Chemical}$) diagram (Fig. 4a) shows different characters from both the GB adsorption ($\Gamma_{Ga}$) diagram and the GB structural width ($\delta_{GB}^{Structural}$) diagram (Fig. 4b). Figure 4a shows that the GB chemical width ($\delta_{GB}^{Chemical}$) mostly increases with increasing Ga bulk composition. In an example illustrated in Fig. 4c, the computed $\delta_{GB}^{Chemical}$ increases from 0.45 to 1.12 nm when the normalized bulk composition ($X/X_{Max}$) increases from 7% to 84% at 500 K. On the other hand, the computed GB structural width ($\delta_{GB}^{Structural}$) diagram (Fig. 4b) shows similar characters as the computed GB excess structural disorder ($\Gamma_{Disorder}$) diagram (Fig. 3b), but with a narrower region of increased $\delta_{GB}^{Structural}$ near the upper half of the bulk solidus line (above $\sim 0.5T_m$). The GB structural width ($\delta_{GB}^{Structural}$) is more strongly correlated with increasing temperature. For instance, the computed $\delta_{GB}^{Disorder}$ increases from 0.66 nm at 300 K to 1.35 nm at 800 K at the same $X/X_{Max}$ of ~24%.

Although GB structural and chemical width diagrams show some distinct characters, these two GB properties are also coupled and correlated. For example, when the temperature rises, the increase of GB disordering not only results in the increase of GB structural width but also enhances Ga segregation and broadens the GB chemical width. In a positive feedback loop, the increased Ga segregation can further prompt GB disordering to broaden the GB structural width. It is worthy to note that when the GB structural width is significantly greater than its chemical width at high temperatures near the upper half of the solidus line (as illustrated, e.g., by the GB # 2 at 800 K in Fig. 4d), this high-temperature GB disordering and widening region is likely akin to the formation of premelting like intergranular films in Ni–S[26], W–Ni[49,50], Mo–Ni[51], and Cu–Zr[52]. Future studies are needed to investigate (and confirm) this premelting like GB disordering region in Al–Ga at high temperatures. On the other hand, the low-temperature region where the GB structural width is smaller than its chemical width suggests the formation of more ordered GB segregation structures, which is consistent with the STEM image for the Al–Ga GB formed at room temperature (Fig. 2). See Supplementary Fig. 2 for further comparison and elaboration.

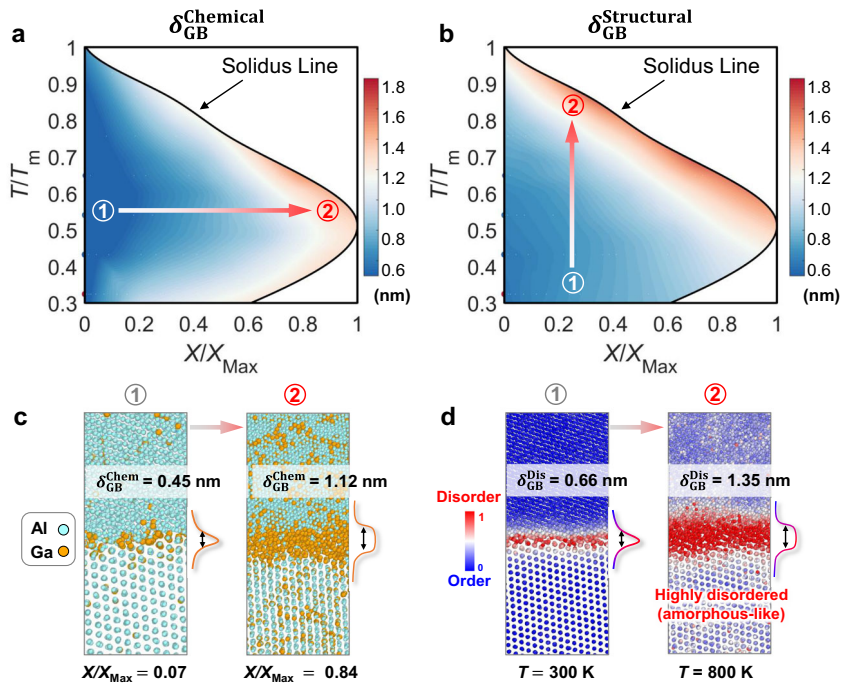

**Fig. 4 Computed diagrams of GB chemical and structural widths.** The corresponding computed GB diagrams of the **a** effective GB chemical width ($\delta_{GB}^{Chemical}$) and **b** effective GB structural (disorder) width ($\delta_{GB}^{Structural}$). **c** Two representative MC/MD-simulated GB structures with different $\delta_{GB}^{Chemical}$ values for $X/X_{Max} = 0.07$ and 0.84, respectively, at 500 K. **d** The color maps of disorder parameter ($\eta_{Dis}$) of two representative MC/MD-simulated GB structures with different $\delta_{GB}^{Structural}$ values for $T = 300$ and 800 K, respectively, at $X/X_{Max} = 0.24$. See Supplementary Fig. 2 for further comparison and illustration.

Based on MC/MD-simulated GB adsorption $\Gamma_{Ga}$ as a function of chemical potential difference and the Gibbs adsorption theory, we also estimated GB energy reduction (due to Ga adsorption): $\Delta\gamma \equiv \gamma - \gamma_0$, where $\gamma$ is the GB energy of an Al–Ga alloy and $\gamma_0$ is the GB energy of pure Al at the same temperature. The corresponding $\Delta\gamma$ diagram, along with the calculation procedure, is documented in Supplementary Fig. 3 and its caption. The $\Delta\gamma$ diagram shows more reduction in the GB energy with Ga adsorption at lower temperatures.

### GB diagrams of simulated mechanical properties

To understand the mechanical properties of the GBs, we performed MD tensile tests on the equilibrium GB structures obtained by hybrid MC/MD simulations. Representative stress ($\sigma$) vs. strain ($\epsilon$) curves for the MD tensile tests for a series of GBs equilibrated at different $\Delta\mu$'s are shown in Supplementary Fig. 4. The maximal tensile stress in the MD simulation is denoted as the MD ultimate strength ($\sigma_{UTS}^{MD}$). In addition, the MD tensile toughness is defined as the integral of the stress: $U_T^{MD} = \int_0^{0.2} \sigma(\epsilon)d\epsilon$. The details of the simulation and analysis procedures are described in the "Methods" section. Subsequently, we computed $\sigma_{UTS}^{MD}$ and $U_T^{MD}$ as functions of temperature and bulk composition to construct GB diagrams of mechanical properties, which are plotted in Fig. 5a and c. It is important to note the MD tensile tests were conducted at a high strain rate of $\sim 5.4 \times 10^8 \text{ s}^{-1}$ due to the limitation of MD simulations. Thus, the absolute values of the MD simulated strength and tensile toughness cannot be directly compared with realistic experimental values, but the relative trends in the GB diagrams of MD mechanical properties can be useful.

The computed GB MD tensile strength ($\sigma_{UTS}^{MD}$) diagram (Fig. 5a) shows that the ultimate strength can be reduced by increasing either the bulk Ga fraction or temperature. Specifically, the detrimental effects of Ga on $\sigma_{UTS}^{MD}$ are more severe at low temperatures (below $\sim 0.4T_M$). In the intermediate temperature region ($\sim 0.4$–$0.6T_M$), a high level of Ga can be tolerated in grains (but with little Ga segregation at GBs according to Fig. 3a). The high-strength region has a half-balloon shape in Fig. 5a. Figure 5b further shows the $\sigma_{UTS}^{MD}$ vs. $X/X_{Max}$ curves at different temperatures. For example, the computed $\sigma_{UTS}^{MD}$ is ~2.95 GPa at 300 K for "clean" GB (with little Ga adsorption) and it is reduced to ~ 1.06 GPa when $X/X_{Max}$ is ~0.70 (near the solubility limit at 300 K). By increasing the temperature, the computed $\sigma_{UTS}^{MD}$ of the "clean" GB can also be reduced to ~1.03 GPa at ~$0.93T_m$ due to the high-temperature reduction of the strength. Again, the high absolute values of the MD strengths are related to the unrealistic high strain rate of the MD tensile tests; thus, the comparison of relative strengths is more useful.

The computed GB MD tensile toughness ($U_T^{MD}$) diagram (Fig. 5c) can illustrate more details about GB fracture behaviors. Notably, a ductile–brittle transition (DBT) line can be identified in Fig. 5c. The GB become brittle even with a small amount of Ga addition at low temperatures, but a higher bulk Ga fraction is needed to induce a DBT at a higher temperature (Fig. 5c). When temperature increases to ~$0.7T_m$ or higher, the GB becomes almost fully ductile, but the computed $U_T^{MD}$ is still reduced due to the high-temperature softening. The corresponding $\sigma_{UTS}^{MD}$ vs. $X/X_{Max}$ curves at various temperatures (Fig. 5d) clearly shows that a low level of Ga segregation at GB can trigger DBT at a low temperature, while GB can tolerate more Ga segregation with ductile failure at a higher temperature.

### GB composition–structure–property relationships

It is interesting to examine the GB composition–structure–property relationships with computed GB diagrams. By comparing

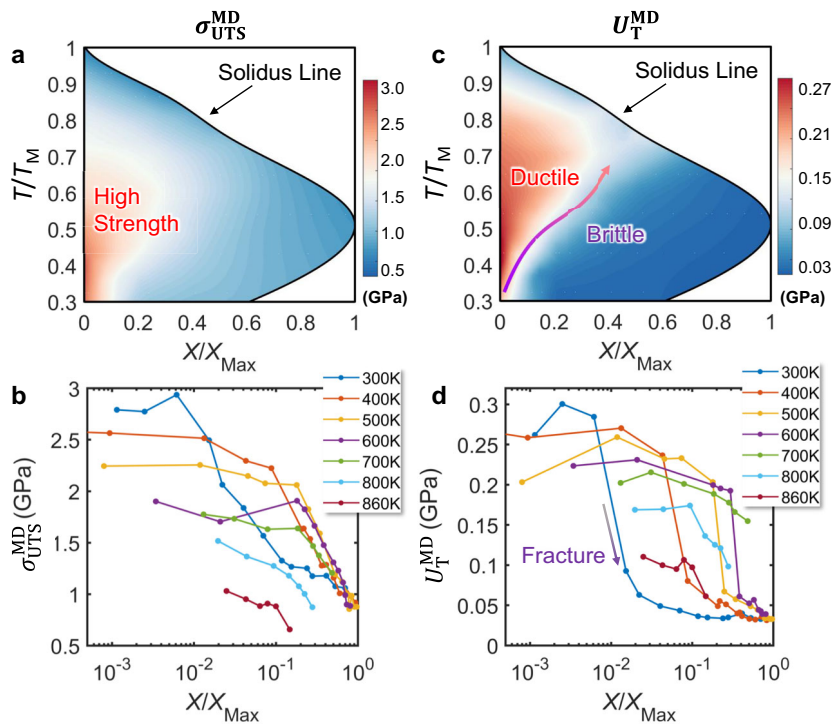

**Fig. 5 Computed GB diagrams of MD-simulated mechanical properties.** Computed GB diagrams of (**a**) MD ultimate tensile strength ($\sigma_{UTS}^{MD}$) and (**b**) MD tensile toughness ($U_T^{MD}$) of the Ga-doped Al asymmetric Σ81 GB from the MD tensile tests based on the hybrid MC/MD simulated equilibrium GB structures. The computed (**c**) $\sigma_{UTS}^{MD}$ and (**d**) $U_T^{MD}$ vs. $X/X_{Max}$ curves at various temperatures.

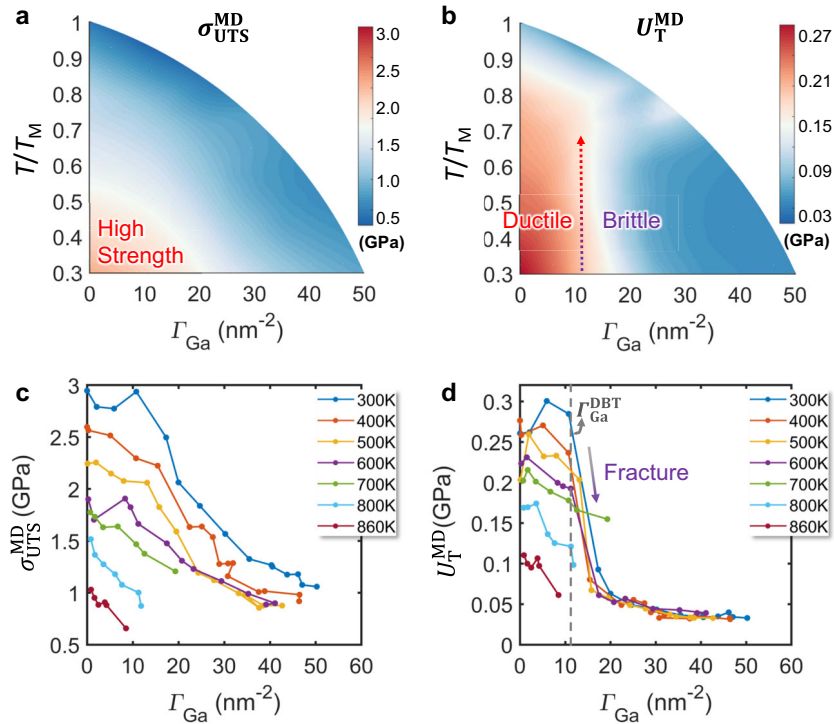

**Fig. 6 Analysis of GB mechanical properties vs. GB adsorption at various temperatures.** The computed GB **a** $\sigma_{UTS}^{MD}$ and **b** $U_T^{MD}$ diagrams were plotted as functions of normalized temperature ($T/T_M$) and GB adsorption ($\Gamma_{Ga}$). The dashed line in panel **b** indicates the approximate position of the ductile-to-brittle transition (DBT). The computed **c** $\sigma_{UTS}^{MD}$ and **d** $U_T^{MD}$ vs. GB adsorption ($\Gamma_{Ga}$) curves at various temperatures. The dashed line in panel **d** indicates the critical threshold of GB adsorption ($\Gamma_{Ga}^{DBT}$) for intergranular fracture.

Fig. 5a and c with Fig. 3a, it is evident that the two computed mechanical properties are inversely correlated with GB adsorption ($\Gamma_{Ga}$). To further investigate these segregation–property relationships, we plot the GB $\sigma_{UTS}^{MD}$ and $U_T^{MD}$ diagrams as functions of temperature and $\Gamma_{Ga}$ in Fig. 6a and b. Figure 6c shows that the computed $\sigma_{UTS}^{MD}$ always decays with increasing $\Gamma_{Ga}$ at all temperatures, while the magnitude of $\sigma_{UTS}^{MD}$ vs. $\Gamma_{Ga}$ curve decreases with increasing temperature uniformly. In addition, Fig. 6d shows that the computed $U_T^{MD}$ curves always decay with increasing the $\Gamma_{Ga}$, which is consistent with the well-known effect of Ga segregation on embrittlement. Interestingly, the computed $U_T^{MD}$ values suddenly drop when $\Gamma_{Ga}$ reaches a critical value. Notably, this GB DBT occurs at the almost identical value of $\Gamma_{Ga} \approx 11$ nm$^{-2}$ when the temperature is <700 K (as indicated by the dashed line in Fig. 6d). This is also plotted as a vertical line (up to ~ $0.7 T_m$) in the computed GB $U_T^{MD}$ diagram (Fig. 6b). Here, we denote this critical value of $\Gamma_{Ga}$ for the DBT as $\Gamma_{Ga}^{DBT}$, which will be used in a phenomenological model presented in the next section.

It is interesting to note that for the current case of Al–Ga, the DBT is mostly determined (only) by the GB adsorption $\Gamma_{Ga}$ (at $T < \sim 0.7 T_m$), while other GB structural characters ($\Gamma_{Disorder}$, $\delta_{GB}^{Chemical}$, and $\delta_{GB}^{Structural}$) have fewer influences. This may be a rather unique character of the Al–Ga system as a simple system with strong GBE but significant solid solubility of Ga in Al (that suppresses GB structural transitions). We hypothesize that other GB structural characters can have more significant impacts on mechanical properties in some other systems with limited solid solubility and the high tendency for GB structural transitions, such as Ni–Bi[13,28,53], Ni–S[26], W–Ni[49,50], or Mo–Ni[16,51].

Comparing Fig. 5b (MD tensile toughness $U_T^{MD}$ diagram) with Supplementary Fig. 3 (GB energy reduction $\Delta\gamma$ diagram), we found that the DBT line can also be drawn based on $\Delta\gamma$ (roughly at the position that $|\Delta\gamma|$ is 50% of its maximum value). This can be explained from the same underlying physics of drawing DBT line based on GB adsorption ($\Gamma_{Ga}$), since $\Delta\gamma$ is obtained based on an integral of $\Gamma_{Ga}$.

Beyond GBE and DBT, the computed GB width diagrams (Fig. 4) can provide additional useful information to predict trends in liquid metal corrosion (GB penetration) and LME behaviors, which supplements the GB adsorption and disorder diagrams that only represent the GB excess quantities. The widening of effective interfacial structural and chemical thickness near (but below) the bulk solidus line can have significant implications in the fast (invasive) Ga penetration at GBs in LME. The computed GB width diagrams (Fig. 4a, b) suggest such interfacial widening is primarily promoted by Ga adsorption (analogues to a case of prewetting[54]) at $T < \sim 0.5 T_m$, but resulted more by interfacial disordering (akin to premelting[55]) at $T > \sim 0.5 T_m$. In general, the computed GB diagrams suggest the formation of nanoscale GB precursor (coupled prewetting and premelting) films in front of the intergranular penetration of thick (bulk) liquid Ga at Al GBs during LME or liquid metal corrosion processes[38,41].

### Mechanism of GB DBT

To understand the mechanism of segregation-induced GB DBT, we further analyzed the GB structural evolution during the MD tensile tests. Figure 7 shows three GB structures and their common neighbor analysis (CNA) at the initial stage (i.e., strain $\epsilon = 0\%$; Fig. 7a–c) and deformed stage ($\epsilon = \sim 6.5\%$; Fig. 7d–f) of the MD tensile tests at 300 K. These GB structures include an unsegregated GB ($\Gamma_{Ga} = 0$ nm$^{-2}$), a weakly segregated GB ($\Gamma_{Ga} = 6.0$ nm$^{-2}$; before the DBT), and a strongly segregated GB ($\Gamma_{Ga} = 17.2$ nm$^{-2}$; right after the DBT), as shown in Fig. 7a–c. The CNA shows that the face-centered cubic (FCC) structures (colored in green in Fig. 7a–c) dominate with high fractions of ~93–94% for all un-strained GB structures, but the hexagonal close-packed (HCP) structures (colored in red in Fig. 7d and e) form in the deformed GB structures, especially for clean and weakly segregated GBs (Fig. 7d, e). Here, the formation of HCP-like structures in FCC alloys can be ascribed to the formation of stacking faults from the GB regions[56]. The stacking fault can form via the deformation-induced emission of Shockley partial dislocation with a Burger vector $\mathbf{b} = \frac{1}{6}\langle 112\rangle$. Since the dislocations contribute to the ductility of metals[57,58], the ductile behavior of clean and weakly segregated GBs can be ascribed to the generation of partial dislocations. However, in the strongly segregated GB (above DBT) with a critical strain that is about to induce fracture (Fig. 7f), we cannot observe any HCP-like structure in the bulk region.

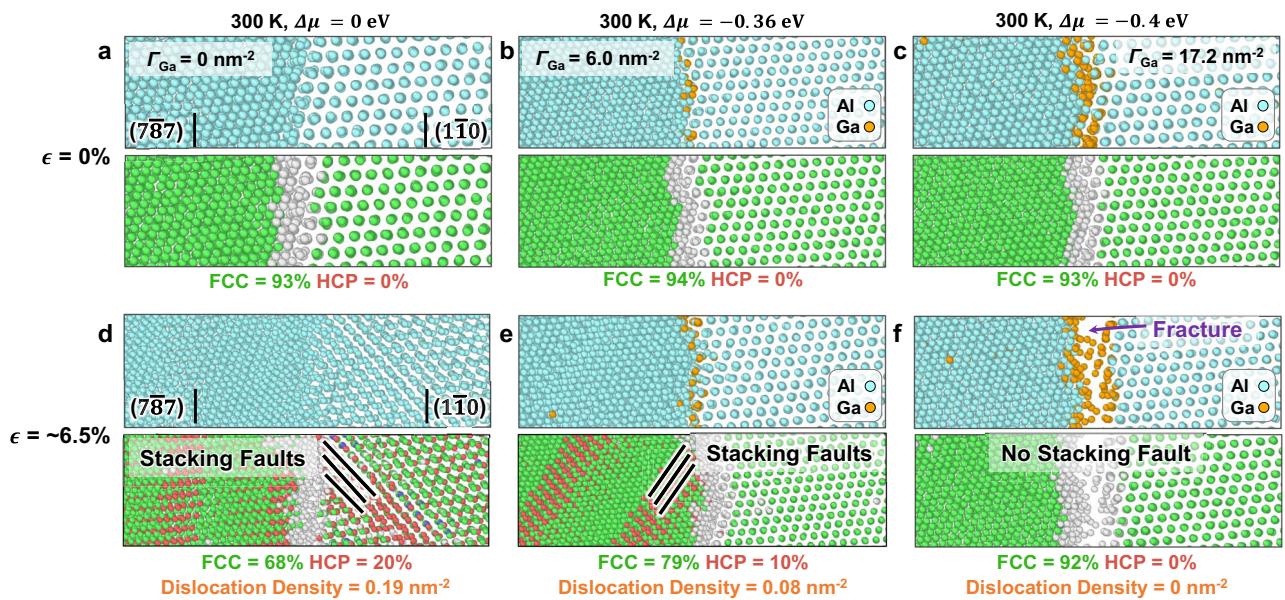

**Fig. 7 GB structural evolution during MD tensile tests.** MC/MD-simulated GB structures and common neighbor analysis (CNA) of **a** an unsegregated GB ($\Gamma_{Ga} = 0\ nm^{-2}$), **b** a weakly segregated GB ($\Gamma_{Ga} = 6.0\ nm^{-2}$), and **c** a strongly segregated GB ($\Gamma_{Ga} = 17.2\ nm^{-2}$), respectively, at the initial stage of the MD tensile test (strain $\epsilon = 0\%$) at $T = 300\ K$. Deformed GB structures and CNA of **d** the unsegregated GB ($\Gamma_{Ga} = 0\ nm^{-2}$), **e** the weakly segregated GB ($\Gamma_{Ga} = 6.0\ nm^{-2}$), and **f** the strongly segregated GB ($\Gamma_{Ga} = 17.2\ nm^{-2}$), respectively, at the strain $\epsilon = \sim 6.5\%$ at $T = 300\ K$. This strain value was selected because it is the minimal strain to induce GB ductile–brittle transition based on stress–strain curve shown in Supplementary Fig. 4. The atoms colored in green in CNA represent face-centered cubic (FCC) structure, and those colored in red represent hexagonal close-packed (HCP) structure.

Next, we performed a detailed analysis of the segregation effect on the formation of stacking faults and dislocations with deformation. By examining CNA maps shown in Fig. 7d–f, we found that the faction of the HCP-like structure decreases from 20% in the clean GB to 10% in the weakly segregated and 0% in the strongly segregated GB. Since HCP-like structure results from the formation of stacking faults, this indicates that increasing the Ga segregation can gradually reduce the formation of stacking faults from the GB regions. This blocking effect is further supported by a dislocation analysis, where the density of Shockley partial dislocation decreases from 0.19 to 0.08 to 0 $nm^{-2}$ with increasing Ga segregation at the GB (Fig. 7d–f). Thus, it appears that the strong Ga segregation can reduce the formation of dislocations and stacking faults from the GB to induce GB DBT when the GB excess of Ga exceeds a critical value (i.e., $\Gamma_{Ga}^{DBT}$ in Fig. 6b).

We note that the reduced dislocation emission and stacking fault formation may result from a reduction in the energy of separation (decohesion) with strong Ga segregation. In other words, the brittle intergranular fracture can take place with strong Ga segregation, before the emission of partial dislocation and formation of stacking faults.

### A phenomenological model
Following the Fowler–Guggenheim isotherm for surface adsorption that was adopted by Hondros and Seah for GB segregation[59],
we can express the relation between GB adsorption ($\Gamma_{Ga}$) and bulk Ga fraction ($X$) in an approximation as

$$\frac{\Gamma_{Ga}}{\Gamma_{Ga}^o - \Gamma_{Ga}} = \frac{X}{1-X} \exp\left[-\frac{\Delta h_{Seg}^o + \alpha(\Gamma_{Ga}/\Gamma_{Ga}^o)}{k_B T}\right] \quad (1)$$

where $\Gamma_{Ga}^o$ represents the effective total number of the Ga adsorption sites per unit area at the GB, $\Delta h_{Seg}^o$ is the intrinsic segregation enthalpy (in the dilute limit), $\alpha$ is a parameter to represent the interaction between the Ga adsorbates (or Al–Ga pair-interaction), and $k_B$ is the Boltzmann constant. Equation (1) can be reduced to the famous Langmuir–Mclean model[60] for $\alpha = 0$ eV. We should note that Eq. (1) only represents a simplified site-occupying model (as an approximation). Nonetheless, we can fit the hybrid MC/MD simulated GB adsorption $\Gamma_{Ga}$ values (represented in Fig. 3a) with Eq. (1) and obtain the following parameters:

$$\begin{cases} \Gamma_{Ga}^o = 48.03\ nm^{-2} \\ \Delta h_{Seg}^o = -0.113\ eV \\ \alpha = -0.025\ eV \end{cases} \quad (2)$$

Here, $\Gamma_{Ga}^o$ can be also considered as the "saturation" level of Ga segregation at the GB. The fitted value agrees with the maximal value of $\sim 50\ nm^{-2}$ in the computed GB adsorption diagram (Fig. 3a), but it is substantially above the one monolayer ($\Gamma_{Ga}^{ML} = 8.6\ nm^{-2}$ for the $(1\bar{1}0)$ plane). The negative value of $\Delta h_{Seg}^o$ indicates that Ga atoms are preferable to segregate at Al GBs. Furthermore, the negative value of $\alpha$ suggests that increasing $\Gamma_{Ga}$ can lower the segregation enthalpy of Ga and therefore prompt segregation. A negative $\alpha$ also suppresses the first-order GB adsorption transition.

Figure 8a shows that the parity plot of the $\Gamma_{Ga}$ values predicted by the fitted analytical model vs. MC/MD simulations, which agree well with each other. The calculated root-mean-square error (RMSE) between analytical model prediction and hybrid MC/MD simulations is $\sim 3.8\ nm^{-2}$ ($\sim 7.6\%$ of $\Gamma_{Ga}^o$). Furthermore, the GB adsorption diagram predicted by the analytical model (Fig. 8b) shows similarity to the hybrid MC/MD simualted GB diagram (Fig. 3a) overall, albeit subtle differences in the DBT transition region and near the solidus line. While it can capture the general trends in Ga adsorption, we should acknowledge that this simple phenomenological model does not consider the effects of significant Ga–Al mixing and interfacial disorder and cannot predict other GB structural features.

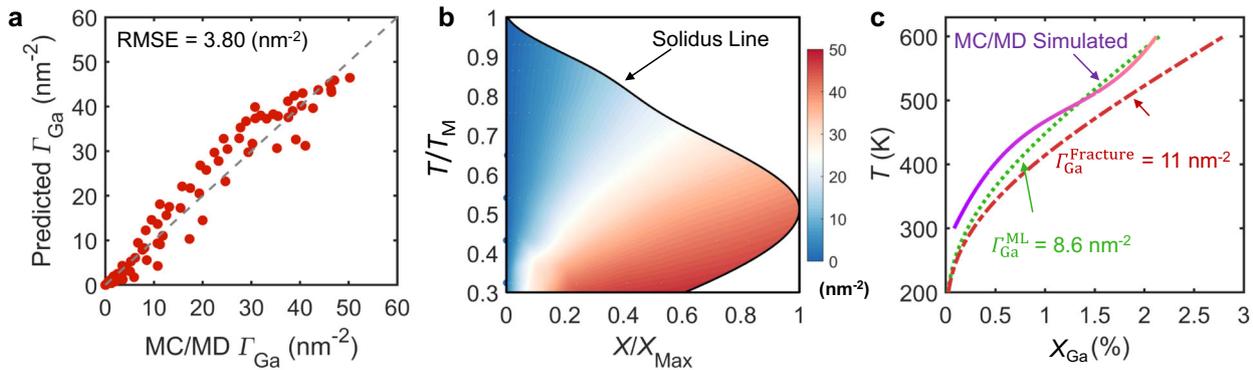

**Fig. 8 A phenomenological model for fitting GB adsorption and predicting the ductile-to-brittle transition (DBT) line. a** Parity plot of model-predicted vs. hybrid MC/MD simulated GB adsorption ($\Gamma_{Ga}$). **b** Model-predicted GB adsorption ($\Gamma_{Ga}$) diagram as a function of $T/T_m$ and $X/X_{Max}$. **c** The ductile-to-brittle transition lines plotted in the $T$-$X_{Ga}$ space from the hybrid MC/MD simulations vs. the model predictions using the critical Ga adsorption values of $\Gamma_{Ga} = \Gamma_{Ga}^{DBT}$ (the critical value from MD tensile tests) and $\Gamma_{Ga} = \Gamma_{Ga}^{ML}$ (monolayer).

Based on the analytical model and fitted parameters, we can predict the DBT line. By adopting $\Gamma_{Ga}^{DBT} = 11$ nm$^{-2}$ (the critical value from MD tensile simulations), we calculated the corresponding bulk composition vs. temperature for the DBT line and plotted it as the red dashed line in Fig. 8c, which agree well with hybrid MC/MD simulated DBT line (the purple color in Fig. 8c). Alternatively, we can simply adopt one monolayer adsorption as the DBT threshold: $\Gamma_{Ga}^{DBT} = \Gamma_{Ga}^{ML} \approx 8.6$ nm$^{-2}$ (estimated based on the plane density of the lower-index ($1\bar{1}0$) plane of this $\Sigma 81$ GB). The predicted DBT line (the green dotted line in Fig. 5c) is even closer to the hybrid MC/MD simulated DBT line.

While a simplified analytical model can predict the DBT with reasonable accuracies (Fig. 8), the hybrid MC/MD simualted GB diagrams (Figs. 3–6) can still reveal more details of the GB segregation, structural, and mechanical properties.

## DISCUSSION
The computed GB diagram (Fig. 3a) shows the GB excess of Ga can be as high as ~50 nm$^{-2}$, which is about 5.6 Ga monolayers. This suggests the occurrence of Ga multilayer adsorption at the Al general GB. Interestingly, the DBT still occurs at approximately one monolayer level (Fig. 6b). A Fowler–Guggenheim type model with a high saturation level (~5.6 monolayers of pure Ga, which can result in an even wider interfacial width with Al–Ga mixing at the GB) and a moderate negative pair-interaction parameter $\alpha$ can fit general trends in the GB adsorption and predict the DBT line reasonably well. The negative fitted value also suggests the Al–Ga interaction at the GB is preferred over Al–Al and Ga–Ga, which is consistent with the absence of first-order adsorption transition in the computed GB diagram (Fig. 3a).

The general GB adsorption behavior of Al–Ga is in contrast to the other classical GBE and LME system, Ni–Bi[13]. Here, Ga has a large solid solubility in the Al-based FCC phase. Thus, the GB segregation can easily go into the multilayer region and follow partial order of (i.e., partially occupy the Al sites of) the abutting Al grains, with more disordered segregation at the GB core (Fig. 2); this proposed "graded order/disorder/order" structure of the GB segregation region at low temperatures (e.g., 300 K) is schematically illustrated in Supplementary Fig. 2. The adsorption transition in the Al–Ga system is continuous with a negative $\alpha$ (i.e., Ga segregation can be more favorable with increasing $\Gamma_{Ga}$). In contrast, Bi has limited solid solubility in the Ni-based FCC phase. Consequently, the Ni general GBs exhibit Bi bilayer adsorption (i.e., one monolayer on each abutting grain, with little Ni–Bi mixing), which can further undergo reconstruction to form highly ordered interfacial superstructures[28,53]. In contrast in Al–Ga ($\alpha < 0$), a first-order GB adsorption transition occurs in Ni–Bi with a positive pair-interaction parameter (i.e., $\alpha > 0$) as suggested by an Ising-type lattice model[13] (but a reliable interatomic potential does not exist for the Ni–Bi system to enable atomistic simulations).

While the GB adsorption is continuous in the current case of Al–Ga, first-order GB adsorption transitions can occur in systems with positive $\alpha$ values (e.g., Ni–Bi[13]), which represent one of a spectrum of possible first-order GB transitions[1,2,10]. In such cases, computed GB diagrams (with well-defined first-order GB transitions, e.g., those for Ni–Bi[13], Mo–Ni[22], and Si–Au[23] systems) can be considered as the GB counterparts to bulk phase diagrams. Interestingly, our simulations showed that an abrupt DBT (i.e., a "first-order" like transition in the GB mechanical property occurring at the critical $\Gamma_{Ga} \approx \Gamma_{Ga}^{DBT}$) can occur in the Ga-doped Al alloy (Fig. 5c) even without a first-order GB transition thermodynamically (Fig. 3).

Our MD tensile tests suggest that strong Ga segregation (above the critical $\Gamma_{Ga}^{DBT}$) can reduce the emission of dislocation and formation of stacking faults from the GB and promote brittle intergranular fracture instead. In a prior study[42], Zhou et al. suggested that Ga segregation may facilitate the emission of dislocations with a pre-existing crack. The apparently different behaviors can be ascribed to the different simulation settings. Zhou et al.'s prior work used an atomistic model with a crack, but we conducted MD tensile testing on the equilibrium GB structure based on a bicrystal geometry without a crack.

Notably, this study constructed GB diagrams of computed mechanical properties, including MD-predicted $U_T^{MD}$ and $\sigma_{UTS}^{MD}$. While the specific values predicted from MD tensile tests are expected to differ from experiments due to the high strain rate of the MD tensile simulations, the relative trends, e.g., the DBT and the relatively high-strength region in GB diagrams of computed mechanical properties can offer useful information and prediction. In general, GB diagrams of various other properties can also be constructed via conducting additional simulations based on the temperature- and composition-dependent equilibrium GB structures, thereby opening numerous opportunities.

In summary, using a general GB in a classic GBE/LME system, Ga-doped Al, as our model system, we have extended the computation of GB diagrams from adsorption and interfacial disorder to include GB structural and chemical widths to better represent the structural details of the GB segregation underpinning LME. Most notably, GB diagrams have been further extended to include computed mechanical properties to establish and investigate the GB composition–structure–property relationships. Temperature- and composition-dependent GB diagrams can be computed for a spectrum of interfacial thermodynamic, structural, mechanical, and potential other functional properties. This work points to a direction to develop computed GB properties diagrams (beyond the basic thermodynamic characters) that can be generally useful in materials design and materials science, with potentially broad impacts.

## METHODS
### Sample preparation and STEM experiments
We prepared transmission electron micrography (TEM) samples of the Al general GBs with Ga segregation following a well-documented procedure[45]. First, we punched the TEM samples with a 3 mm disc from a high-purity (99.99%) Al polycrystalline thin foil (~5 μm in grain size and 100 μm in thickness) by a Gatan disc puncher. The thickness of discs was reduced to ~25 μm by a Gatan disc grinder. The pre-thinned disc was scratched near the center to expose fresh surfaces. Then, we quickly transferred it onto a hot plate at 110 °C and placed a small high-purity (99.999%) Ga particle on the scratched surface to allow the penetration of liquid Ga. The residual oxide particles on the surface were picked by a tweezer to avoid strains in the TEM sample. The sample was sandwiched between two copper rings by glue (G1 glue, Gatan) to prevent hazardous Ga contamination. The reinforced Al disc was thinned by a precision ion polishing system (PIPS; Gatan Model 695) until a hole was drilled by the ion beam. A Thermo Fisher Scientific Talos TEM and a Themis Z AC TEM were used to characterize the atomic-level interfacial structures of randomly selected Al GBs with Ga segregation.

### Hybrid MC/MD simulations
The GB structures for the simulations were constructed by GBstudio website[61] and Pymatgen code[62,63] based on coincidence-site lattice (CSL) theory. Two types of GBs, i.e., a symmetric tilt Σ11 (113)//(113) GB and an asymmetric Σ81 (1$\bar{1}$0)//($\bar{7}$87) GB of mixed tilt and twist characters (to represent a general GB) were considered in this work. The symmetric Σ11 GB was built based on the STEM image reported in a prior work[33], which contains 18,480 atoms in the unit cell. The asymmetric Σ81 GB was constructed to mimic an experimentally observed GB randomly selected from a polycrystalline sample in this work. Here, we adopted the low-index (1$\bar{1}$0) surface terminal plane observed in the randomly selected general GB from the experiment. Then, we found a matching high-index plane with a Miller index of ($\bar{7}$87) to form a CSL with a large Σ value of 81 to satisfy a required periodic boundary condition while better representing a general GB. All subsequent simulations were based on a large unit cell containing 19,440 atoms for the asymmetric Σ81 GB to represent a general GB.

All atomistic simulations were performed using LAMMPS code[64]. An Al–Ga EAM potential originally developed by the Srolovitz group[44] was adopted for all simulations. To test this EAM potential, the Al–Ga bulk phase diagram (Supplementary Fig. 1) was simulated, which agreed well with the experimentally measured phase diagram[65]. The simulated Al melting temperature is 925 K (close to the experimental value of 933.5 K with <1% relative error). We note that the calculated maximum solid solubility is slightly higher than experimental value[65]. Such small discrepancies are typical for EAM potentials. We represent our simulation results in the normalized temperature (with respect to the Al melting temperature) and normalized Ga fraction (with respect to the maximum solid solubility). Overall, the agreement suggests the EAM potential is robust and reliable for the Al–Ga system.

The initial GB structure was first relaxed at a high temperature of 700 K for 500 ps with a time step of 0.1 fs by classical MD simulations in constant NPT ensembles. Next, the hybrid Monte Carlo and molecular dynamics (hybrid MC/MD) simulations in isobaric semi-grand canonical (constant N(Δμ)PT) ensembles were carried out to dope Ga atoms into Al GBs at a given temperature and a fixed chemical potential difference $\Delta\mu = \mu_{Al} - \mu_{Ga}$, where $\mu_{Al}$ and $\mu_{Ga}$ is the chemical potential of Al and Ga, respectively. Five MC trial moves were conducted between each MD step with a 0.1 fs MD time step. One million hybrid MC/MD steps were performed for each simulation to achieve convergence.

### Computing GB properties and GB diagrams
To construct GB diagrams as a function of temperature (T) and the bulk Ga fraction (X), the hybrid MC/MD simulations were performed from 300 to 860 K with a temperature step of 100 K. At each temperature, a series of MC/MD simulations with increasing Δμ were conducted until the bulk composition of Ga ($X_{Ga}$) reaches the solid solubility limits. The chemical potential difference Δμ will be converted to bulk Ga fraction based on the hybrid MC/MD simulated equilibrium bulk compositions.

The centrosymmetric parameter (CSP) was calculated by summing the pairs of opposite neighbors of a specific lattice, where the CSP of the ideal crystal CSP is zero due to cancellation of opposite neighbor pairs but crystal defect will have positive CSP values[47]. The atomic structures shown in this work were visualized by the OVITO code[48].

Several GB thermodynamic and structural properties were computed to construct GB diagrams.

First, the GB excess of Ga can be determined by

$$\Gamma_{Ga} = \int_{-\infty}^{+\infty} [X_{Ga}(z) - X_{Ga}(\pm\infty)]dz,$$

where $X_{Ga}(z)$ is the 2D averaged Ga fraction profile (see, e.g., Fig. 2f) and $X_{Ga}(\pm\infty) \equiv X$ is the bulk Ga fraction.

Second, the GB disorder parameter was calculated for each atom following a procedure proposed by Chua et al. ($\eta_{Dis} = 0$ for a perfect crystal and $\eta_{Dis} = 1$ for a perfect liquid)[46].

Furthermore, we calculated two types of GB effective widths based on the chemical and disorder profiles. Specifically, the GB chemical and structural widths were, respectively, calculated by measuring the full widths at the half maximum (FWHMs) of 2D averaged Ga compositional profile $X_{Ga}(z)$ and structural disorder profile $\eta_{Dis}(z)$ along the z-direction. In both cases, we used the same sampling number of 150 and step width of ~0.12 nm to calculate $X_{Ga}(z)$ and $\eta_{Dis}(z)$ profiles across the GB to allow a fair comparison of the width (to exclude artifacts caused by different coarse-graining schemes).

### STEM simulation
STEM images were simulated by using the QSTEM program. The hybrid MC/MD simulated Ga-doped Al GB structures were used as the input atomic structures. The thickness of the simulated sample was set to be 3.5 nm (as the thin specimen limit). The scattering semi-angle for HAADF imaging was set to be 70 mrad, the convergence angle was adopted as 15 mrad. The spherical aberration coefficient was set to be 0 μm.

### MD tensile tests for mechanical properties
Using the hybrid MC/MD simulated equilibrium GB structures, a series of uniaxial tensile tests of GBs was carried out by isothermal–isobaric (constant NPT) MD simulations at desirable temperatures. The strain rate was set to ~5.4 × 10$^8$/s and tensile tests were performed until the final strain reaches ~0.2. See Supplementary Fig. 3 for representative stress–strain curves at different Δμ values (with different $\Gamma_{Ga}$) at 300 K. The maximal tensile stress based on the MC/MD simulations were considered as the MD ultimate strength ($\sigma_{UTS}^{MD}$). The MD tensile toughness ($U_T^{MD}$) was calculated by integrating the stress–strain curve (for the strain from 0 to 0.2) based on MD simulations. The common neighbor analysis (CNA) and dislocation analysis (DXA) of the GB structures during MD tensile tests were performed by the OVITO code[48].

## DATA AVAILABILITY
The data that support the findings of this study are available from the corresponding author upon reasonable request.

## CODE AVAILABILITY
The code used to calculate the results of this study are available from the corresponding author upon reasonable request.

## REFERENCES

1. Cantwell, P. R. et al. Grain boundary complexions. *Acta Mater.* **62**, 1–48 (2014).
2. Cantwell, P. R. et al. Grain boundary complexion transitions. *Annu. Rev. Mater. Res.* **50**, 465–492 (2020).
3. Lejček, P., Šob, M. & Paidar, V. Interfacial segregation and grain boundary embrittlement: an overview and critical assessment of experimental data and calculated results. *Prog. Mater. Sci.* **87**, 83–139 (2017).
4. Raabe, D. et al. Grain boundary segregation engineering in metallic alloys: a pathway to the design of interfaces. *Curr. Opin. Solid State Mater. Sci.* **18**, 253–261 (2014).
5. Luo, J. Stabilization of nanoscale quasi-liquid interfacial films in inorganic materials: a review and critical assessment. *Crit. Rev. Solid State Mater. Sci.* **32**, 67–109 (2007).
6. Sutton, A. P. & Balluffi, R. W. *Interfaces in Crystalline Materials* (Oxford Scientific Publications, 1995).
7. Hart, E. W. Two-dimensional phase transformation in grain boundaries. *Scr. Metall.* **2**, 179–182 (1968).
8. Dillon, S. J., Tang, M., Carter, W. C. & Harmer, M. P. Complexion: a new concept for kinetic engineering in materials science. *Acta Mater.* **55**, 6208–6218 (2007).
9. Tang, M., Carter, W. C. & Cannon, R. M. Diffuse interface model for structural transitions of grain boundaries. *Phys. Rev. B* **73**, 024102 (2006).
10. Kaplan, W. D., Chatain, D., Wynblatt, P. & Carter, W. C. A review of wetting versus adsorption, complexions, and related phenomena: the rosetta stone of wetting. *J. Mater. Sci.* **48**, 5681–5717 (2013).
11. Luo, J. Developing interfacial phase diagrams for applications in activated sintering and beyond: current status and future directions. *J. Am. Ceram. Soc.* **95**, 2358–2371 (2012).
12. Luo, J. Liquid-like interface complexion: from activated sintering to grain boundary diagrams. *Curr. Opin. Solid State Mater. Sci.* **12**, 81–88 (2008).
13. Zhou, N., Yu, Z., Zhang, Y., Harmer, M. P. & Luo, J. Calculation and validation of a grain boundary complexion diagram for Bi-doped Ni. *Scr. Mater.* **130**, 165–169 (2017).
14. Nie, J., Chan, J. M., Qin, M., Zhou, N. & Luo, J. Liquid-like grain boundary complexion and sub-eutectic activated sintering in CuO-doped $TiO_2$. *Acta Mater.* **130**, 329–338 (2017).
15. Luo, J. & Shi, X. M. Grain boundary disordering in binary alloys. *Appl. Phys. Lett.* **92**, 101901 (2008).
16. Shi, X. & Luo, J. Developing grain boundary diagrams as a materials science tool: a case study of nickel-doped molybdenum. *Phys. Rev. B* **84**, 014105 (2011).
17. Zhou, N. & Luo, J. Developing grain boundary diagrams for multicomponent alloys. *Acta Mater.* **91**, 202–216 (2015).
18. Wang, L. & Kamachali, R. D. Density-based grain boundary phase diagrams: application to Fe–Mn–Cr, Fe–Mn–Ni, Fe–Mn–Co, Fe–Cr–Ni and Fe–Cr–Co alloy systems. *Acta Mater.* **207**, 116668 (2021).
19. Mishin, Y., Boettinger, W. J., Warren, J. A. & McFadden, G. B. Thermodynamics of grain boundary premelting in alloys. I. Phase-field modeling. *Acta Mater.* **57**, 3771–3785 (2009).
20. Johansson, S. A. E. & Wahnstrom, G. First-principles study of an interfacial phase diagram in the V-doped WC–Co system. *Phys. Rev. B* **86**, 035403 (2012).
21. Hu, C., Zuo, Y., Chen, C., Ong, S. P. & Luo, J. Genetic algorithm-guided deep learning of grain boundary diagrams: addressing the challenge of five degrees of freedom. *Mater. Today* **38**, 49–57 (2020).
22. Yang, S., Zhou, N., Zheng, H., Ong, S. P. & Luo, J. First-order interfacial transformations with a critical point: breaking the symmetry at a symmetric tilt grain boundary. *Phys. Rev. Lett.* **120**, 085702 (2018).
23. Hu, C. & Luo, J. First-order grain boundary transformations in Au-doped Si: hybrid Monte carlo and molecular dynamics simulations verified by first-principles calculations. *Scr. Mater.* **158**, 11–15 (2019).
24. Williams, P. L. & Mishin, Y. Thermodynamics of grain boundary premelting in alloys. II. Atomistic simulation. *Acta Mater.* **57**, 3786–3794 (2009).
25. Yamaguchi, M., Shiga, M. & Kaburaki, H. Grain boundary decohesion by impurity segregation in a nickel–sulfur system. *Science* **307**, 393–397 (2005).
26. Hu, T., Yang, S., Zhou, N., Zhang, Y. & Luo, J. Role of disordered bipolar complexions on the sulfur embrittlement of nickel general grain boundaries. *Nat. Commun.* **9**, 2764 (2018).
27. Dillon, S. J., Tai, K. & Chen, S. The importance of grain boundary complexions in affecting physical properties of polycrystals. *Curr. Opin. Solid State Mater. Sci.* **20**, 324–335 (2016).
28. Luo, J., Cheng, H., Asl, K. M., Kiely, C. J. & Harmer, M. P. The role of a bilayer interfacial phase on liquid metal embrittlement. *Science* **333**, 1730–1733 (2011).
29. Schweinfest, R., Paxton, A. T. & Finnis, M. W. Bismuth embrittlement of copper is an atomic size effect. *Nature* **432**, 1008–1011 (2004).
30. Duscher, G., Chisholm, M. F., Alber, U. & Rühle, M. Bismuth-induced embrittlement of copper grain boundaries. *Nat. Mater.* **3**, 621–626 (2004).
31. Rice, J. R. & Wang, J.-S. Embrittlement of interfaces by solute segregation. *Mater. Sci. Eng. A* **107**, 23 (1989).
32. Senel, E., Walmsley, J. C., Diplas, S. & Nisancioglu, K. Liquid metal embrittlement of aluminium by segregation of trace element gallium. *Corros. Sci.* **85**, 167–173 (2014).
33. Sigle, W., Richter, G., Rühle, M. & Schmidt, S. Insight into the atomic-scale mechanism of liquid metal embrittlement. *Appl. Phys. Lett.* **89**, 121911 (2006).
34. Nam, H.-S. & Srolovitz, D. J. Effect of material properties on liquid metal embrittlement in the Al–Ga system. *Acta Mater.* **57**, 1546–1553 (2009).
35. Joseph, B., Picat, M. & Barbier, F. Liquid metal embrittlement: a state-of-the-art appraisal. *Eur. Phys. J. Appl. Phys.* **5**, 19–31 (1999).
36. Kolman, D. G. A review of recent advances in the understanding of liquid metal embrittlement. *Corrosion* **75**, 42–57 (2019).
37. Nicholas, M. G. & Old, C. F. Liquid-metal embrittlement. *J. Mater. Sci.* **14**, 1–18 (1979).
38. Ludwig, W., Pereiro-López, E. & Bellet, D. In situ investigation of liquid Ga penetration in Al bicrystal grain boundaries: grain boundary wetting or liquid metal embrittlement? *Acta Mater.* **53**, 151–162 (2005).
39. Rajagopalan, M., Bhatia, M. A., Tschopp, M. A., Srolovitz, D. J. & Solanki, K. N. Atomic-scale analysis of liquid-gallium embrittlement of aluminum grain boundaries. *Acta Mater.* **73**, 312–325 (2014).
40. Nam, H.-S. & Srolovitz, D. J. Molecular dynamics simulation of Ga penetration along grain boundaries in Al: a dislocation climb mechanism. *Phys. Rev. Lett.* **99**, 025501 (2007).
41. Ludwig, W. & Bellet, D. Penetration of liquid gallium into the grain boundaries of aluminium: A synchrotron radiation microtomographic investigation. *Mater. Sci. Eng. A* **281**, 198–203 (2000).
42. Zhou, G. et al. Liquid metal embrittlement mechanism. *Sci. China Technol. Sci.* **42**, 200–206 (1999).
43. Shen, M. et al. The interfacial structure underpinning the Al–Ga liquid metal embrittlement: disorder vs. order gradients. *Scr. Mater.* **204**, 114149 (2021).
44. Nam, H.-S. & Srolovitz, D. J. Molecular dynamics simulation of Ga penetration along Σ5 symmetric tilt grain boundaries in an Al bicrystal. *Phys. Rev. B* **76**, 184114 (2007).
45. Hugo, R. C. & Hoagland, R. G. In-situ TEM observation of aluminum embrittlement by liquid gallium. *Scr. Mater.* **38**, 523–529 (1998).
46. Chua, A. L.-S., Benedek, N. A., Chen, L., Finnis, M. W. & Sutton, A. P. A genetic algorithm for predicting the structures of interfaces in multicomponent systems. *Nat. Mater.* **9**, 418–422 (2010).
47. Kelchner, C. L., Plimpton, S. J. & Hamilton, J. C. Dislocation nucleation and defect structure during surface indentation. *Phys. Rev. B* **58**, 11085–11088 (1998).
48. Stukowski, A. Visualization and analysis of atomistic simulation data with ovito—the open visualization tool. *Model. Simul. Mater. Sci. Eng.* **18**, 015012 (2009).
49. Gupta, V. K., Yoon, D.-H., Meyer, H. M. & Luo, J. Thin intergranular films and solid-state activated sintering in nickel-doped tungsten. *Acta Mater.* **55**, 3131–3142 (2007).
50. Luo, J., Gupta, V. K., Yoon, D. H. & Meyer, H. M. Segregation-induced grain boundary premelting in nickel-doped tungsten. *Appl. Phys. Lett.* **87**, 231902 (2005).
51. Shi, X. & Luo, J. Grain boundary wetting and prewetting in Ni-doped Mo. *Appl. Phys. Lett.* **94**, 251908 (2009).
52. Khalajhedayati, A., Pan, Z. & Rupert, T. J. Manipulating the interfacial structure of nanomaterials to achieve a unique combination of strength and ductility. *Nat. Commun.* **7**, 10802 (2016).
53. Yu, Z. et al. Segregation-induced ordered superstructures at general grain boundaries in a nickel–bismuth alloy. *Science* **358**, 97–101 (2017).
54. Cahn, J. W. Critical point wetting. *J. Chem. Phys.* **66**, 3667–3672 (1977).
55. Dash, J. G., Rempel, A. M. & Wettlaufer, J. S. The physics of premelted ice and its geophysical consequences. *Rev. Mod. Phys.* **78**, 695–741 (2006).
56. Hu, C., Medlin, D. L. & Dingreville, R. Disconnection-mediated transition in segregation structures at twin boundaries. *J. Phys. Chem. Lett.* **12**, 6875–6882 (2021).
57. He, B. B. et al. High dislocation density–induced large ductility in deformed and partitioned steels. *Science* **357**, 1029–1032 (2017).
58. Wu, Z. X., Zhang, Y. W. & Srolovitz, D. J. Dislocation–twin interaction mechanisms for ultrahigh strength and ductility in nanotwinned metals. *Acta Mater.* **57**, 4508–4518 (2009).
59. Hondros, E. D. & Seah, M. P. The theory of grain boundary segregation in terms of surface adsorption analogues. *Metall. Trans. A* **8**, 1363–1371 (1977).
60. Wynblatt, P. & Chatain, D. Anisotropy of segregation at grain boundaries and surfaces. *Metall. Mater. Trans. A* **37**, 2595–2620 (2006).

npj

C. Hu et al.

61. Ogawa, H. GBstudio: a builder software on periodic models of CSL boundaries for molecular simulation. *Mater. Trans.* **47**, 2706–2710 (2006).
62. Ong, S. P. et al. Python materials genomics (pymatgen): a robust, open-source python library for materials analysis. *Comput. Mater. Sci.* **68**, 314–319 (2013).
63. Zheng, H. et al. Grain boundary properties of elemental metals. *Acta Mater.* **186**, 40–49 (2020).
64. Plimpton, S. Fast parallel algorithms for short-range molecular dynamics. *J. Comput. Phys.* **117**, 1–19 (1995).
65. Okamoto, H. & Massalski, T. *Binary Alloy Phase Diagrams* (ASM International, 1990).

## ACKNOWLEDGEMENTS
The calculations were performed at the Triton Shared Computing Cluster (TSCC) at the UCSD. We thank Prof. Kiran Solanki from Arizona State University for providing the code file for the EAM potential originally developed by Prof. Srolovitz's group.

## AUTHOR CONTRIBUTIONS
J.L. conceived the idea and supervised the work. C.H. performed the simulations. Y.L. and Z.Y. conducted STEM experiments. C.H. and J.L. wrote the manuscript. All co-authors reviewed and approved the manuscript.

## COMPETING INTERESTS
The authors declare no competing interests.

## ADDITIONAL INFORMATION
**Supplementary information** The online version contains supplementary material available at https://doi.org/10.1038/s41524-021-00625-2.

**Correspondence** and requests for materials should be addressed to Jian Luo.

**Reprints and permission information** is available at http://www.nature.com/reprints

**Publisher's note** Springer Nature remains neutral with regard to jurisdictional claims in published maps and institutional affiliations.